\documentclass[twocolumn,10pt]{asme2ej}

\usepackage{epsfig} 
\usepackage{url}            
\usepackage{hyperref}       

\title{Using Deep Neural Networks for Estimating Loop Unrolling Factor}

\author{Asma Balamane
    \affiliation{
     \textit{ LMCS (ESI)} \\
      Algiers, Algeria \\
      ea\_balamane@esi.dz\\
    }	
}
\author{Zina Taklit
    \affiliation{
      \textit{LMCS (ESI)} \\
      Algiers, Algeria \\
      ez\_taklit@esi.dz\\
    }	
}
\author{Riyadh Baghdadi
    \affiliation{
      CSAIL (MIT)\\
      Massachusetts, USA \\
      baghdadi@mit.edu
    }	
}

\begin{document}
\maketitle    
\begin{abstract}
\textbf{\textit{Abstract}\rule{1em}{0.4pt}To achieve high performance in modern processors, compilers should optimize programs. We address in this paper Loop Unrolling optimization, proposing a novel approach based on deep neural networks to automatically optimize loops in TIRAMISU. \\TIRAMISU is a new language to create a code of high performance. It allows to separate between the algorithm and its optimizations\cite{tiramisu}.}\\ \\
\end{abstract}
\textbf{\textit{Key words}} \hspace{0.2cm} Loop Unrolling, Deep Neural Network, Feature Extraction, TIRAMISU compiler.
\section{Introduction}
\par Optimizing programs requires deep expertise. On one hand, it is a tedious task, because it requires a lot of tests to find out the best combination of optimizations to apply with their best factors. On the other hand, this task is critical, because it may degrade the program’ performance instead of improving it. The automatization of this task can deal with this problem and permit to obtain good results.
Optimizing loops that take the most significant part of the program execution time plays a crucial role to achieve best performance.
 In this paper, we address Loop unrolling optimization, by proposing a deep Neural Network model to predict the optimal unrolling factor for TIRAMISU’s programs.
 \par TIRAMISU is a polyhedral framework \cite{PRADELLE2011polyedrique} designed to generate high performance code for multiple platforms including multicores, GPUs, and distributed machines. TIRAMISU introduces a scheduling language with novel commands to explicitly manage the complexities that arise when targeting these systems\cite{tiramisu}.
\section{Loop unrolling}
\paragraph*{}
Loop unrolling is the transformation in which the loop body is replicated “k” times where “k” is a given unrolling factor. It is used to reduce overhead by decreasing the number of iterations and hence the number of branch operations. Loop unrolling enables other optimizations, many of which target the memory system. The main important advantage of loop unrolling is that it exposes instruction level parallelism (ILP) to the compiler\cite{supervised2005}. Unrolling improves performance almost in all cases where it is applied in a significant way \cite{David1994Compilertransformations}. However, if the loop unrolling is not carefully applied, it may negatively affect other important optimizations and reduce overall performance. Choosing the right factor of unrolling is also very important. The best unrolling factor reduces execution time and improves global performance. Therefore, through this research, we aim to design a model that predict the best unrolling factor.
\section{Learning Loop Unrolling Factor Models}
\paragraph*{}In this section, we present our model design that predicts the best loop unrolling factor. We use TIRAMISU compiler \cite{tiramisu} as execution platform, a polyhedral compiler that allows flexible application of different loop optimizations.
\label{LUF}
\subsection{Input program features}
\paragraph*{} For any machine learning technique, selecting the best input features is a crucial step. Since our contribution focus on the loop unrolling optimization which is a local optimization, we implement a method to extract features automatically for each loop nest (TIRAMISU computation). This loop nests abstraction summarizes the characteristics that influence the execution efficiency on modern processors and gives a high-level representation independently on the execution architecture platforms.  \\
The features vector contains the most important nest loop characteristics such as the number of loop nest levels, dependencies between loop nest levels and the loop operations characteristic, etc.. In the other hand, other loop optimizations can be already applied on the loop nests, we associate for each loop nest the features list of schedule optimizations.
The table \ref{table_ASME} presents a subset of features given by the automatic extraction method.
\begin{table}[t]
\caption{Figure and table captions do not end with a period}

\label{table_ASME}
\begin{tabular}{l}

\hline
 \textbf{Loop levels characteristics}\\
\hline
Number of the loop nest levels  \\
Number of dependencies between loop nest levels  \\
List of dependent levels for each loop nest level \\
Loop span for each level \\
Whether there is a predicate before each loop level \\
  \hline 
\textbf{Operations characteristics}  \\
\hline
Operation loop level / operation rank in the loop level  \\
Number of variables/invariants used in the operation  \\
Operation histogram per operands type  \\
Loads/stores histogram per operands type  \\
Number of library calls for each loop nest level  \\
  \hline 
\textbf{Schedule (Optimizations) characteristics}  \\
   \hline
  
  Whether the optimization is applied  \\
  The loop nest levels the optimization is applied  \\
  Factors used for each applied optimization \\
  List of dependent loop nest (global optimizations case)\\ 
\hline
\end{tabular}
\end{table}
\paragraph*{} Collecting training data must be carefully done. We use \href{https://github.com/Tiramisu-Compiler/tiramisu/tree/master/utils/code_generator}{\textbf{TIRAMISU code generator tool}} to generate parameterized unrolled code with different other optimizations cases. We exhaustively search for the optimal unrolling factor of the generated programs to create the training data set.

\subsection{Using Deep Neural Networks (DNN) to Learn Loop Unrolling factor model}
 \paragraph*{}
We used a deep neural network to construct a supervised classification model. It allows the prediction of the best unrolling factor. In a classification model, outputs (classes) are predefined. It receives as input  a set of labeled learning data to learn how to classify new programs. The architecture of the neural network is based on the typical architecture of multilayer perceptron (MLP)\cite{GARDNER19982627}. The model must predict the output among the defined classes that represent the range of possible values of the unrolling factor.\\
\begin{figure}[h!]
\includegraphics[scale=0.6]{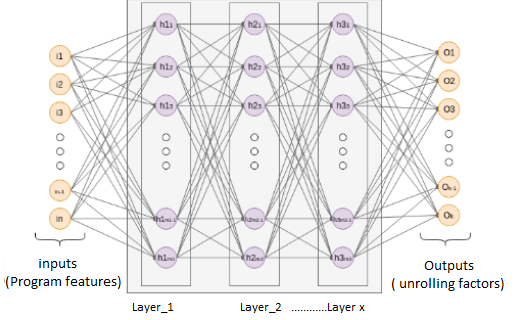}
\caption{DNN model for estimating the best unrolling factor.}
\label{fig:DNN}
\end{figure}
\paragraph*{}
We adopted an empirical strategy to define the hyperparameters of the DNN. We found that four hidden layers with 500, 400, 250 and 100 neurons in each layer respectively, gives the best accuracy. For each layer we dichotomically tested the cases of neurons number.
Concerning the other model hyperparameters, we summarized tests results in the table \ref{tab:HyperParamsTable}
\begin{table}[t]
  \centering
  \caption{subset of the model hyperparameters.}

\begin{tabular}{|c|c| }
\hline 
 \textbf{Activation function} & ReLu                  \\  \hline 
 \textbf{Optimization algorithm} & ADAM\cite{AdamOptimizer}    \\  \hline
\textbf{ Learning rate  } &  $10^{-3}$         \\  \hline
 \textbf{Initialisation algorithm} & Random\_uniform  \\  \hline
 \textbf{Number of iterations}  & Early stopping technique   \\  \hline
\end{tabular}

\label{tab:HyperParamsTable}
  \end{table}
\section{Experimental results} 
The system is evaluated thanks to a set of benchmarks\footnote{find benchmarks implementation in \url{https://github.com/AsmaBALAMANE/tiramisu/tree/master/benchmarks/Automatic_unroll}}. We have first implemented the exhaustive exploration of unrolling factors technique. It represents a reference for comparing the prediction model results. For each benchmark, we have launched an exhaustive exploration of the unrolling factors in order to define the best factor and to compare it with the factor predicted by the implemented  model. We consider for each benchmark three test cases (depending on the data size or the applied optimizations). \\
for each test case we evaluate the \textit{PC} and  \textit{SP} metrics that represent ratios between programs execution time with the optimal and predicted unrolling factors and without applying unrolling optimization  
\begin{center} \textbf{PC} = $ optimal\_exec / predit\_exec$\end{center}
\begin{center} \textbf{SP} = $without\_exec / predit\_exec$ \end{center} 
   
\begin{figure}[h]
	\begin{center}
		\includegraphics[scale=0.62]{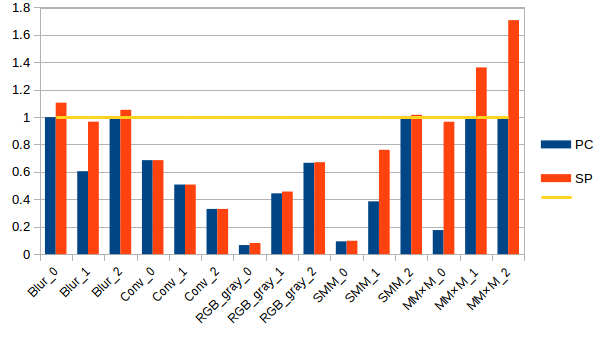}
	\end{center}
	\caption{Test results with the different benchmarks test cases.}
	\label{fig:test_synth}
\end{figure}
We note in Figure\ref{fig:test_synth} that PC and SP rates vary from one benchmark to another. In fact, for the $MM\times M$ and Blur benchmark, we recorded fairly positive rates. \\This means that the model is able to learn good predictions for dealing with the data locality problem that is required in both benchmarks.To synthesize, the model learns high\-level features from low level features. It gives predictions of the best unrolling factor for new programs with a precision of up to 20\%. This shows that the model learns and it exceeds the random prediction (whose precision is 14\%)
\section{Future work}
\paragraph*{} 
The accuracy of our model is influenced by the lack of data. Thereby, we are generating more data to improve model accuracy.
\\In this paper, we presented the general design of based Deep learning predictor for unrolling optimization. In order to develop a complete automatic optimization method for TIRAMISU compiler, the final stage of our work will refine the presented method by adding the loop optimizations selection method.

\bibliographystyle{asmems4}

\bibliography{asme2e}

\end{document}